\documentclass[aip,reprint]{revtex4-1}
\usepackage{amsmath,amsbsy}

\renewcommand\vec[1]{\boldsymbol{#1}}
\def\be{\begin{equation}}
\def\ee{\end{equation}}

\begin{document}
\title{Completing the Li\'enard-Wiechert potentials: The origin of the delta function fields for a charged particle in hyperbolic motion}
\affiliation{Physics Department, Haverford College, Haverford, PA, 19014, USA}
\author{Daniel J. Cross}
\email{dcross@haverford.edu}
\date{\today}
\begin{abstract}
Calculating the electromagnetic fields of a uniformly accelerated charged particle is a surprisingly subtle problem that has been long discussed in the literature.  In particular, the fields calculated from the Li\'enard-Wiechert potentials fail to satisfy Maxwell's equations.  While the correct fields have been obtained many times and through various means, it has remained unclear why the standard approach fails.  We identify and amend the faulty step in the Li\'enard-Wiechert construction and provide a new direct calculation of the fields and potentials for a charge in hyperbolic motion.
\end{abstract}

\maketitle
\section{Introduction}\label{sec:intro}

The Li\'enard-Wiechert (LW) construction yields an explicit expression for the electromagnetic fields of a charged particle in arbitrary motion.\cite{G,J}  However, it has been pointed out\cite{FG,B} that in at least one instance, namely for a particle undergoing relativistic hyperbolic motion (constant proper acceleration), this ``standard formula'' fails: the resulting fields do not satisfy the Maxwell equations on all of spacetime as they lack certain delta function terms.  While the missing terms have been reconstructed in several different ways,\cite{FG,B} these approaches involve amending or supplementing the hyperbolic motion in some way, and they do not explain why hyperbolic motion causes the standard construction to fail.  We address this question here and find that physically the problem is that the particle's speed approaches $c$ in the infinite past, while mathematically the problem is  handling the delta function that defines the retarded time in that limit. 
We begin in Sec.~\ref{sec:review} with a review of the LW construction of the electromagnetic potentials.  In Sec.~\ref{sec:origin} we directly produce the missing electromagnetic field terms through a slight alteration of the standard construction.  Finally, in Sec.~\ref{sec:completing} we explain the fault in the LW construction, amend it, and produce the missing potential terms.

\section{Review of the Electromagnetic Potentials}\label{sec:review}
The electromagnetic potentials may be expressed as integrals of the charge density and current over all space and time\cite{J}
\begin{align}
\label{eq:V}V(\vec x,t)&=\frac{c}{4\pi\epsilon_0}\int G\rho(\vec x',t')d\vec x'dt'\\
\label{eq:A}\vec A(\vec x,t)&=\frac{1}{4\pi\epsilon_0c}\int G\vec J(\vec x',t')d\vec x'dt',
\end{align}
where $G$ is the (retarded) Green's function, given by\cite{Note1}
\be\label{eq:G1}
G=\frac{\delta(ct-ct'-R)}{R}\Theta(t-t'),
\ee
and where $\vec R=\vec x-\vec x'$ is the relative position vector, and $R=|\vec R|$ is its length.  The Green's function propagates the effects of a point source at $(\vec x',t')$ to all points $(\vec x,t)$ along the forward light-cone, $c(t-t')=R=|\vec x-\vec x'|$, as enforced by the delta function.  A useful equivalent representation of $G$ is\cite{Note2}
\be\label{eq:G2}
G=2\delta(\tau^2)\Theta(t-t'),\;\;\tau^2=c^2(t-t')^2-R^2.
\ee

For a point charge $q$ following the path $\vec\xi(t)$, the charge density is $\rho(\vec x',t')=q\delta[\vec x'-\vec \xi(t')]$, and the current density is $\vec J=\rho\dot{\vec\xi}= \rho\vec v$.  With these expressions the potentials become
\begin{align}
V&=\frac{qc}{4\pi\epsilon_0}\int G\delta[\vec x'-\vec\xi(t')]d\vec x'dt'\\
\vec A&=\frac{q}{4\pi\epsilon_0c}\int G\vec v\delta[\vec x'-\vec\xi(t')]d\vec x'dt'.
\end{align}
Carrying out the spatial integral using the delta function localizes the Green's function to the particle's worldline, and the potentials simplify to
\be\label{eq:pots}
V=\frac{qc}{4\pi\epsilon_0}\int G dt',\;\;
\vec A=\frac{q}{4\pi\epsilon_0c}\int G\vec vdt',
\ee
where now $\vec R=\vec x-\vec\xi(t')$ in $G$.  Performing the remaining integral (see Sec.~\ref{sec:completing}) over $t'$ yields the Li\'enard-Wiechert potentials
\be\label{eq:LW}
V=\left.\frac{qc}{4\pi\epsilon_0}\cdot 
\frac{1}{cR-\vec R\cdot\vec v}\right|_{t_r},\;\;
\vec  A=\left.\frac{\vec v}{c^2}V\right|_{t_r},
\ee
where the notation indicates that all quantities are to be evaluated at retarded time $t_r$, which is the (unique) solution to $t-t_r-R(t_r)/c=0$ with $t_r<t$, and represents when the past lightcone of $(\vec x,t)$ intersected the charge's worldline.  For a charge in hyperbolic motion along the $z$-axis, the electromagnetic fields $\vec E=-\nabla V-\partial\vec A/\partial t$ and $\vec B=\nabla\times\vec A$ calculated from these potentials will fail to satisfy the Maxwell equations on the $ct+z=0$ plane,\cite{FG} missing a term proportional to $\delta(ct+z)$.

\section{The origin of delta function fields}\label{sec:origin}
Instead of taking derivatives of the completed potentials to obtain the fields, let us instead, following Barut,\cite{Barut} compute the derivatives \emph{before} performing the time integrals in Eq.~(\ref{eq:pots}), e.g.
\be\label{eq:intdV}
\nabla V=\frac{qc}{4\pi\epsilon_0}\int \nabla Gdt'.
\ee
According to Eq.~(\ref{eq:G2}), away from the charge itself ($R\neq0,t\neq t'$), $G$ is a function only of $\tau^2$, so that using the chain rule repeatedly
\be
\nabla G=\frac{dG}{d\tau^2}\nabla\tau^2=\frac{dG}{dt'}\frac{dt'}{d\tau^2}\nabla\tau^2.
\ee
It is straightforward to show that
\be \nabla\tau^2=-2\vec R\textrm{\;\;and\;\;}
\frac{dt'}{d\tau^2}=-\frac12\cdot\frac{1}{c^2(t-t')-\vec v\cdot \vec R}.\ee
Using these expressions the integral in Eq.~(\ref{eq:intdV}) becomes
\be
\int \nabla Gdt'=
\int \left(\frac{\vec R}{c^2(t-t')-\vec v\cdot \vec R}\right) \left(\frac{dG}{dt'}dt'\right),
\ee
which can be integrated by parts to give
\begin{align}\label{eq:parts1}
&\left.\frac{G\vec R}{c^2(t-t')-\vec v\cdot \vec R}\right|_{-\infty}^{\infty}\\
-&\label{eq:parts2}
\int G \frac{d}{dt'}\left[\frac{\vec R}{c^2(t-t')-\vec v\cdot \vec R}\right]dt'.
\end{align}

Thus there are two distinct contributions to $\nabla V$.  The integral in Eq.~(\ref{eq:parts2}) can be evaluated directly,\cite{Barut} yielding an expression identical to that obtained by taking the gradient of the LW scalar potential, Eq.~(\ref{eq:LW}).  Let us therefore label this contribution as $\nabla V^{LW}$.  The other contribution is the boundary term, Eq.~(\ref{eq:parts1}).  Since the boundary is at infinity, let us label this contribution to the gradient as $\nabla V^\infty$.  We are accustomed to having boundary terms at infinity vanish, so may be tempted to dismiss this term without a thought, but let us not be so hasty here and actually evaluate it.  Going back to Eq.~(\ref{eq:G1}), the step function is $G$ is zero unless $t>t'$, so the upper limit $t'\to+\infty$ gives zero, and we can set $\Theta=1$ for evaluating the lower limit $t'\to-\infty$.  Using the delta function we can replace $c(t-t)'$ with $R$ in the denominator, leaving (putting in the zero value of the upper limit explicitly)
\be\label{eq:blim}
0-\lim_{t'\to-\infty}\frac{\vec R\delta(ct-ct'-R)}{cR^2-\vec v\cdot \vec RR}.
\ee

For hyperbolic motion $R\to\infty$ as $t'\to-\infty$, so the argument of the delta function has the indeterminate form $\infty-\infty$.  For hyperbolic motion along the $z$-axis, $z'=\sqrt{b^2+(ct')^2}$, and using polar coordinates $(s,\theta,z)$ as in Ref.~\onlinecite{FG}, we have
\be
R=\sqrt{s^2+(z-z')^2},
\ee
which asymptotically becomes\cite{Note3}
\be
R\to -ct'-z-\frac{s^2+b^2}{2ct'}+O(1/t')^2.
\ee
The argument of the delta function is then
\be\label{eq:asympt}
ct-ct'-R\to ct+z+\frac{s^2+b^2}{2ct'}\to ct+z,
\ee
so that the delta function is supported on the $ct+z=0$ plane, precisely where the missing field term is supposed to be.

Curiously, had the asymptotic speed been less than $c$, this delta function would be off at infinity (not along $ct+z=0$), and this boundary term would contribute nothing to field.  E.g.\ for $z'=(v_\infty/c)\sqrt{b^2+(ct')^2}$, with $v_\infty<c$, then $R\to -z-v_\infty t'$, and
\be
ct-ct'-R\to ct+z+(v_\infty-c)t'\to\infty,
\ee
as $v_\infty-c< 0$.  We may conclude that physically the trouble with the fields for hyperbolic motion is caused by the particle speed asymptotically approaching $c$.\cite{Note4}

The denominator of Eq.~(\ref{eq:blim}) is also indeterminate as $t'\to-\infty$.  Asymptotically the first term is 
\be
cR^2\to c(ct'+z)^2+c(s^2+b^2)+O(1/t').
\ee
To evaluate the second term, $\vec v\cdot \vec RR$, first note that $\vec v\cdot \vec R=(z-z')(dz'/dt')$, and that we can write $dz'/dt'=ct'/z'$.  Then
\be
\vec v\cdot \vec RR\to c(ct'+z)^2+(c/2)(s^2+b^2)+O(1/t').
\ee
When taking the difference the leading terms cancel and $(c/2)(s^2+b^2)$ survives in the limit.  At this point Eq.~(\ref{eq:blim}) reads
\be
\nabla V^\infty=-\frac{q}{2\pi\epsilon_0}\cdot\frac{\delta(ct+z)}{s^2+b^2}\lim_{t'\to-\infty} \vec R.
\ee
For motion along the $z$-axis $s'=0$, so $R_s=s$, and the $s$-component of the electric field is (the vector potential component $A_s=0$ for motion along the $z$ axis)
\be
E^\infty_s=-\nabla_s V^\infty=\frac{q}{2\pi\epsilon_0}
\frac{s}{s^2+b^2}\delta(ct+z),
\ee
which is precisely the delta function field of Ref.~\onlinecite{FG} [last term of their Eq.~(C1); see also Eq.~(III.11) of Ref.~\onlinecite{B}].  

For the $z$-component of the field we need to evaluate $R_z=z-z'$, but $z'\to\infty$ as $t'\to-\infty$.  However, the vector potential $A_z$ also contributes to $E_z$.  Let us evaluate $\partial A_z/\partial t$ following the same procedure as $\nabla V$.  First we need 
\be
\frac{\partial G}{\partial t}=
\frac{dG}{dt'}\frac{dt'}{d\tau^2}\frac{\partial\tau^2}{\partial t}=
-\frac{dG}{dt'}\frac{c^2(t-t')}{c^2(t-t')-\vec v\cdot \vec R}.
\ee
Integrating by parts gives two contributions: the standard $\partial\vec A^{LW}/\partial t$ and the boundary term
\begin{align}
\nonumber\frac{\partial A_z^\infty}{\partial t}&=
\frac{1}{4\pi\epsilon_0c^2}\cdot
\left.\frac{-Gc^2(t-t') v}{c^2(t-t')-\vec v\cdot \vec R}\right|_{-\infty}\\
&=-\frac{q}{2\pi\epsilon_0}\cdot\frac{\delta(ct+z)}{s^2+b^2}\lim_{t'\to-\infty}c(t-t'),
\end{align}
which also blows up as $t'\to-\infty$.  The complete $z$-component of the electric field arising from these boundary terms is
\begin{align}
\nonumber E^\infty_z&=-\nabla_z V^\infty-\frac{\partial A^\infty_z}{\partial t}\\
&=
\nonumber \frac{q\delta(ct+z)}{2\pi\epsilon_0(s^2+b^2)}
\lim_{t'\to-\infty}\Big[(z-z')+c(t-t')\Big]\\
&= 0.
\end{align}
The limit gives zero because $z+ct=0$ on account of the delta function while $-z'-ct'\to 0$ as $t'\to-\infty$ for hyperbolic motion.  Finally, there is also a delta function term $B_\theta^\infty$ missing from the magnetic field [not considered in Ref.~\onlinecite{FG}, but see Eq.~(III.11) of Ref.~\onlinecite{B}], which can be obtained as $\left(\nabla\times \vec A^\infty\right)_\theta=-\partial A^\infty_z/\partial s$ following an analogous procedure.  We find
\be
B_\theta^\infty=-\frac{q}{2\pi\epsilon_0c}
\frac{s}{s^2+b^2}\delta(ct+z)=-E_s^\infty/c,
\ee
in agreement with Ref.~\onlinecite{B}.

Boulware\cite{B} found these missing terms by boosting a static Coulomb field and taking the limit as the boost speed approached $c$, identifying the delta function field as ``the original Lorentz transformed Coulomb field of the charge `before' it began its acceleration.''  The present analysis is congruent with Boulware's assessment as the delta terms were obtained from a boundary contribution at infinity.  We have the rather astounding result that a source infinitely remote in space and time produces non-negligible electromagnetic fields if it is moving at the speed of light (more precisely, if is located at past null infinity\cite{Note5}).  This gives some insight into the failure of the usual procedure: because the source is \emph{at} infinity, it lies beyond the reach of the usual expression for the LW potentials.

\section{Completing the Li\'enard-Wiechert construction}\label{sec:completing}
We have successfully derived the missing delta fields, but the procedure we employed raises a rather vexing question: why does simply reversing the order of differentiation and integration make a difference in the value of the field?  To answer this question, consider the nature of the extra terms: they are due to a source \emph{at} infinity.  Recall from Eq.~(\ref{eq:asympt}) that as $t'\to-\infty$ (and $R\to\infty$), the delta function in $G$ becomes  $\delta(ct-ct'-R)\to\delta(ct+z)$, supported on the $ct+z=0$ plane rather than out at infinity. The behavior of the source at infinity is therefore non-trivial, and care must be taken when evaluating the $t'\to-\infty$ limit.  

Before we evaluate the limit, let us first reveal where the standard construction goes awry.  All the steps in Sec.~\ref{sec:intro} are fine up to and including Eq.~\ref{eq:pots}, which is the integral 
\be
\int G dt'=\int\frac{\delta(ct-ct'-R)}{R}dt'.
\ee
The next step is to integrate out the delta function, defining the retarded time in the process.  But this is not a straightforward procedure as the delta function is a nonlinear function of $t'$, so the following identity\cite{GS} is invoked
\be\label{eq:identity}
\delta[f(t')]=\frac{\delta(t'-t_0)}{|\dot f(t_0)|},
\ee
where $t_0$ is the (assumed unique) root of the nonlinear function $f$, and the derivative $\dot f=df/dt'$ in the denominator must not vanish at $t_0$.  In the present context $f(t')=ct-ct'-R$ and  $t_0=t_r$ is the retarded time.  Use of this identify transforms the integral to
\be\label{eq:lwtrans}
\int\frac{\delta(ct-ct'-R)}{R}dt'=
\int\frac{\delta(t'-t_r)}{R|c+\dot R|_{t_0}}dt',
\ee
so that now the delta function can integrated out in the usual way.  This transforms $t'\to t_r$, and the usual LW potentials, Eq.~(\ref{eq:LW}), result.

The trouble is that for hyperbolic motion this procedure is ill-defined in the  $t'\to-\infty$ limit.  Because the particle asymptotically approaches $z'=-ct'$, for every point on the $ct+z=0$ plane the retarded time is the infinite past $t_r=-\infty$.  In this limit the denominator in Eq.~(\ref{eq:lwtrans}) is ill-behaved as $R\to\infty$ while $c+\dot R\to 0$.  Again, this would not have happened had the speed been less than $c$ in the infinite past, as there there would have been no solution for the retarded time, and the integrand in Eq.~(\ref{eq:lwtrans}) would just go to zero.  The mathematical fault in the standard LW construction is therefore the use of this identity, which fails when the particle's speed approaches $c$ in the infinite past.

Let us amend the standard construction by integrating over the delta function directly (near $t'\to-\infty$), without appealing to Eq.~(\ref{eq:identity}).  Using the asymptotic forms of $R$ and of the delta function argument, the integral can be written as 
\be
\int_{-\infty}\frac{\delta(ct-ct'-R)}{R}dt'\to
\int_{-\infty}\frac{\delta(\alpha+\beta/t')}{-ct'}{dt'},
\ee
where we have defined $\alpha=ct+z$ and $\beta=(s^2+b^2)/2c$ (which are independent of $t'$) for brevity.  By changing variables to $u=-\beta/t'$ (so that $u\to 0^+$ as $t'\to-\infty$) the delta function can be directly integrated
\be\label{eq:dvar}
-\int_0\frac{\delta(\alpha+u)}{cu}{du}=-\lim_{\alpha\to 0}\frac{1}{c\alpha},
\ee
which is singular for $\alpha=ct+z=0$.  We anticipate that this expression is proportional to a delta function in $\alpha$.  The coefficient of this delta function is the value of its integral over all $\alpha$, which we now compute.  Going back to Eq.~(\ref{eq:dvar}) and integrating over $\alpha$ first we find
\be
-\int_0 du\int\frac{\delta(\alpha+u)}{u}{d\alpha}=-\int_0\frac{du}{u}=-\lim_{u\to 0^+}\ln u,\\
\ee
so that upon transforming back from $u$ to $t'$ we obtain 
\begin{align}
\nonumber -\lim_{u\to 0^+}\ln u &=
-\lim_{t'\to-\infty} \ln\frac{s^2+b^2}{-2ct'}\\
\nonumber &=-\lim_{t'\to-\infty} \ln\frac{(s^2+b^2)/b^2}{-2ct'/b^2}\\
&=-\ln\frac{s^2+b^2}{b^2}+\lim_{t'\to-\infty}\ln\frac{-2ct'}{b^2}.
\end{align}
In the second line factors of $b^2$ were inserted to set the scale of the logarithms in the third line.  Putting in the pre-factors we obtain for the asymptotic scalar potential
\be
V^\infty=\frac{q\delta(ct+z)}{4\pi\epsilon_0}\left[-\ln\frac{s^2+b^2}{b^2}
+\lim_{{t'}\to-\infty}\ln\frac{-2ct'}{b^2}\right].
\ee
Except for the logarithmically diverging term this agrees with the scalar potential postulated in Ref.~\onlinecite{FG} [their Eq.~(37)].  The asymptotic vector potential $A_z^\infty$ can be handled in exactly the same way.  With $v_z=dz'/dt'\to-c$ we find
$$A^\infty_z=-\frac{q\delta(ct+z)}{4\pi\epsilon_0}\int_{-\infty}Gdt'
=-V^\infty/c.$$
Again, the finite term in the vector potential matches that postulated in Ref.~\onlinecite{FG}.  There are still the divergent terms, but they are inconsequential as they can be removed by a gauge transformation ($V^\infty\to V^\infty-\partial\Lambda/\partial t$ and $A_z^\infty\to A_z^\infty+\partial\Lambda/\partial z$) with the gauge factor
\be
\Lambda=\frac{q\Theta(ct+z)}{4\pi\epsilon_0c}\ln\frac{-2ct'}{b^2},
\ee
applied prior to completing the limit.

In summary, proper evaluation of the delta function in the LW integral produces two terms
\begin{align}
\nonumber V&=V^{LW}+V^\infty\\
\vec A&=\vec A^{LW}+\vec A^\infty,
\end{align}
the standard LW term for normal particle motions with finite retarded times and the boundary term for asymptotic light-like particle motion with an infinite past retarded time.

While hyperbolic motion is quite simple, the asymptotic approach to light speed in the infinite past has surprising physical implications.  We have found that a charge moving at light speed, though infinitely remote in space and time, produces an electromagnetic field.  The failure of the standard LW construction to account for this source lies in the standard manipulation of the delta function, a procedure which is ill-defined in the required limit.  Boulware\cite{B} was apparently aware of this, noting only in passing that the missing delta fields ``can be calculated directly from the retarded field of the uniformly accelerated charge\ldots if the field is carefully treated as a distribution,'' though he presented no such calculation.

\section*{Acknowledgment}
The author would like to thank David Griffiths for bringing this interesting problem to his attention and for providing helpful comments on this manuscript.

%

\end{document}